# Stochastic resonance in Schmitt trigger and its application towards weak signal detection

Yoonkang Kim[1] and Donghyeok Seo[1,*]

[1] Department of Physics & Astronomy, Seoul National University, Seoul 08826 Korea
[*] The author whose correspondence should be addressed.

**Abstract**—This study explores stochastic resonance (SR) in a Schmitt trigger circuit and its application to weak signal detection. SR, a phenomenon where noise synchronizes with weak signals to enhance detectability, was demonstrated using a custom-designed bi-stable Schmitt trigger system. The circuit's bi-stability was validated through hysteresis curve analysis, confirming its suitability for SR studies. Experimental results revealed SR behavior by analyzing signal-to-noise ratio (SNR) responses to noise amplitude variations. Detection experiments were conducted to determine frequency and amplitude of damping sinusoidal pulses. Frequency detection proved effective, albeit with limitations at low frequencies, while amplitude detection faced challenges due to mathematical complexities. Nonetheless, the study highlights SR's potential for weak signal detection, with proposed enhancements to improve detection accuracy. This work underscores the adaptability of classical SR principles to practical detection systems and suggests future applications in advanced detection technologies, including quantum systems.

## I. INTRODUCTION

For detection of unknown signals, noise is treated as impure blending that should be eliminated. However, in terms of stochastic resonance it is not. Stochastic resonance is a phenomenon in which the weak signal and noise is synchronized. In the presence of stochastic resonance, the output signal has a behavior of transition between two values. So presence of noise is required for stochastic resonance. The noise can be intrinsic with respect to the physical system. Else, it can be generated artificially. Now back to the detection of unknown signals, stochastic resonance is known to have potential in detecting weak signals[1]〜[5],[8]. This is the point our research starts. However, stochastic resonance is known to be naturally occurred with bi-stable system and intrinsic noise in it. So the first step would be a proper implementation of bi-stable system. From further modules, basic and advanced elements of circuit were studied. In this respect, bi-stable system made by circuits, especially mainly using the op-amp is selected. With using the characteristic of op-amp and we implemented Schmitt trigger in terms of feedback, which is mentioned in section-theoretical background. After the implementation, observing stochastic resonance will be the next. Since the main purpose is to see it's possibility for the application in detection, two detection experiments are constructed. The following experimental steps and results will give us deeper insight on circuits, especially Schmitt trigger. Then, mainly on bi-stability of physical systems and what really stochastic resonance is. Eventually, by actually detecting the signals, we hope verification of stochastic resonance's capability in use of detection can be done.

### A. Theoretical analysis of Schmitt trigger

A Schmitt trigger is a comparator circuit with hysteresis, having two different thresholds. The comparator, an op-amp with positive feedback has $V_{out}$ connected to $V_n$, voltage at inverting input. $V_p$ as a voltage of non-inverting input and $A$ as a gain, voltage output can be expressed as $V_{out} = A(V_p - V_n)$. $A$ being sufficiently large even for real op-amp, one may assume the saturation with $V_{out}$ fixed as saturation or bias voltage. By positive feedback, $V_p = CV_{out}$, implying $V_p$ of two fixed values, each for each $V_{out}$. Coefficient $C$ is positive constant, determined by the voltage divider between output and non-inverting input. Therefore, $V_{out}$ and $V_p$ have a same sign, $V_p$ being positive when it's larger than $V_n$, negative the other way. Here, we Assign positive $V_p$ as VUT, negative $V_p$ as VLT, $V_{out}$ as $V_{sat+}$ and $V_{sat-}$. Let's increase $V_n$ from an initial state of $V_{out} = V_{sat+}$ and $V_p$ = VUT. $V_{out}$ stays at $V_{sat+}$ until $V_n$ is smaller than VUT then switch to $V_{sat-}$ as it passes VUT, an upper threshold formed at VUT. Now with $V_{out}$ and $V_p$ at negative value, we decrease $V_n$. $V_{out} = V_{sat-}$ until $V_n$ gets to VLT then switch to $V_{sat+}$ as it passes VLT which is a lower threshold. In this way, the comparator with positive feedback has two output voltage $V_{sat+}$, $V_{sat-}$, two threshold VUT, VLT , constituting the hysteresis.

**B. Basic of stochastic resonance and tendency of signal-to-noise ratio**

Noise, the term originated from the Latin word "nausea" meaning seasickness, is undesirable fluctuation. Ruling out the noise has been essential to minimize the error and conserve the original signal. Contrary to this conventional wisdom, stochastic resonance (SR) is a highly unique phenomenon where noise plays the constructive role. In SR, adding Gaussian white noise to the weak signal enables the detection of the signal. For example, in the bi-stable system or the double-well potential, the weak signal alone cannot shift the state in one well to the other. In the presence of appropriate noise, the transition between two states occurs, with the output signal amplified. The common parameter to discuss the quality of the detection through SR is signal-to-noise ratio, SNR. At low amplitude of the noise, only none or few transitions are made, with a low SNR. Increasing the amplitude of the noise leads to a higher number of transitions, its frequency being a lot closer to that of the original signal, hence a higher SNR can be obtained. However, SNR would decrease above a particular level since the transition is now mainly depending on the noise rather than the signal itself, and its frequency differs arbitrarily. So, the SNR follows a resonance-like behavior with the variable being the noise amplitude, while it is the frequency for a common resonance.

**C. Justification of bi-stability through hysteresis curve**

To study the stochastic resonance, implementing appropriate bi-stable system is needed and we should check whether the constructed single system has two stable states. For this confirmation of the bi-stability, plotting a hysteresis curve is one of the well-known methods. In the bi-stable system there are two states and two different thresholds for each transition of the state. When the transition occurs at one threshold, the state shifts and stays at the other state until the system reaches the other threshold. Because the environment is not changing, the thresholds and the states are fixed. Furthermore, each state must not decay into the other state unless it passes the threshold. This aspect of the bi-stable system would form the hysteresis curve, as shown in Fig.1. The depth of each well will not be completely equal to the threshold at the hysteresis curve, but Fig.1. would be enough to demonstrate the connection between the hysteresis and the bi-stable system. Thus, hysteresis curve of the system implies the bi-stability of it.

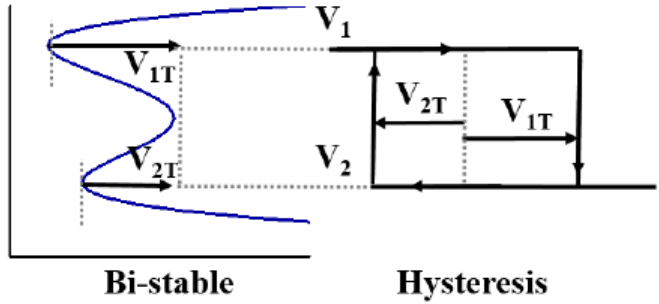

**Fig. 1** Connection between bi-stable system and the hysteresis

**II. Methods**

**A. Implementation of bi-stable system: Schmitt trigger**

Experimental set up is shown in Fig.2. To combine the input signal $V_{input}$ and noise $V_{noise}$, voltage divider with resistance $R = 100\Omega$ was used. Note that this reduces the input signal and noise into half. Combined signal is transmitted to (-) input of Schmitt trigger. Two voltmeter of AD2 were used to measure the combined signal $V_{in}$ and the output signal of Schmitt trigger $V_{out}$. DC power supply of op-amp $V_{dc}$ was set as ±1V resulting the threshold voltage of Schmitt trigger $V_{th} = 0.045$ evaluated by Eq.1. Again, note that the double of $V_{th}$ is required for $V_{in}$ to

$$V_{UT} = \frac{V_{sat} + R_2}{R_1 + R_2}$$
$$V_{LT} = \frac{V_{sat} - R_2}{R_1 + R_2} \text{ [7]} \quad (1)$$

make the transition in $V_{out}$ due to the voltage divider. Since the first goal of our experiment is implementation of bi-stable system with Schmitt trigger, it is important to check whether the Schmitt trigger has bi-stability and thus shows stochastic resonance. At first, the transition of output pulse and SNR-σ curve were observed to verify whether stochastic resonance occurs. It is possible to check whether the stochastic resonance has occurred in the system by inspecting the SNR-σ curve, because the stochastic resonance gives a curve with local maximum. Then, the Hysteresis loop can be obtained to check the bi-stability of Schmitt trigger system. This step is required. Because even though it is obvious that bi-stable systems show stochastic resonance, the inversion isn't. Following measurements were done with LABVIEW program.

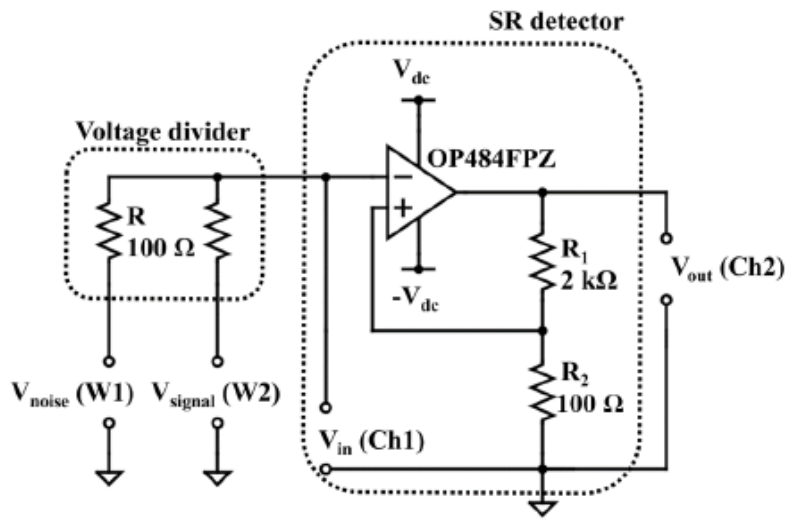

**Fig. 2** Schematic of implemented Schmitt trigger

For observing transition of output signal, with acquisition time and sample rate of 0.4s and 20000Hz were set, respectively. SD of Gaussian white noise was set as 0.05V and the input signal's frequency and amplitude were set as 100Hz and 0.1V. To get SNR, sine wave with amplitude 0.05V, frequency 500Hz was used as $V_{signal}$. Acquisition time and sample rate were set as 0.4s and 20000Hz, respectively. SD of Gaussian white noise $V_{noise}$ varied from 0.01V to 0.20V by increments of 0.005V. For each values of SD, $V_{out}$ was measured. Since FFT analysis gives a relative contribution of each frequency to the subject wave, FFT was done on $V_{out}$ to get SNR. SNR means the relative intensity of input signal compared to the noise, it can be gained by subtracting the mean values of total FFT results from the FFT value at the frequency of input signal (which is 500Hz in this case). All calculation was done in logarithm scale(dB). This gives the SNR-σ curve. Even though the SNR-σ curve shows stochastic character, we can't sure that the system has bi-stability. So further inspection on hysteresis of circuit was done. DC signal was used as $V_{signal}$ and there was no $V_{noise}$ in this case. Acquisition time and sample rate were set as 0.4s and 20000Hz respectively. $V_{out}$ was measured for the two cases; $V_{in}$ increases from -0.2V to 0.2V and $V_{in}$ decreases from 0.2V to -0.2V. From this result, $V_{in} - V_{out}$ curve was plotted to show the hysteresis of Schmitt trigger circuit and to find the real threshold. Additionally, $V_{th}$ dependence of $V_{dc}$ was evaluated for further insight of the threshold of Schmitt trigger system. For fixed DC signal $V_{signal} = 1$, electric potential at (+) input of op-amp was measured. Acquisition time and sample rate were set as 0.1s and 20000Hz respectively. Absolute value of electric potential $V_{th}$ was plotted as a function of $V_{dc}$, and $R^2$ of linear regression was inspected to see the linearity.

**B. Weak signal detection: frequency**

Since the transition rate of $V_{out}$ follows the frequency of $V_{in}$, it is possible to figure out the frequency of input signal by inspecting $V_{out}$. If the input signal is too weak that it can't exceed the threshold of Schmitt trigger, aid of additional noise is needed. In the case of sinusoidal input, amplitude of signal doesn't change so we can find one proper SD of noise as we did in A.2. Now combined signal $V_{in}$ would make transition of $V_{out}$, and it leads to the peak of FFT at the frequency of input signal $V_{signal}$. However, things get more complicated if the amplitude of signal changes with time. To see that still Schmitt trigger can be used as a signal detector, exponentially decaying sine signal which has the form of $V_{signal} = Ae^{-bt}sin2\pi f t$ was tested. We expected that if the rate of decay is too fast compared to the frequency of sine, only few transitions occur for at the beginning of the signal, so the accuracy of measured frequency would be reduced. To check this, input signal was set as $A = 0.1$V, $b = 5$s$^{-1}$ with frequency

$f = 10, 50, 100, 500, 1000, 2000$ Hz. SD of noise was 0.01V, $V_{dc} = 1$V and acquisition time and sample rate were set as 0.4s and 20000Hz respectively. FFT was done on $V_{out}$. Since the FFT has highest peak at $f = 0$, frequency of second highest peak was inspected and compared to that of $V_{signal}$ to calculate error rate. After then, same analysis was done for different noise SD 0.02V and 0.03V to see that the how the result changes depending on the noise.

**C. Weak signal detection: amplitude**

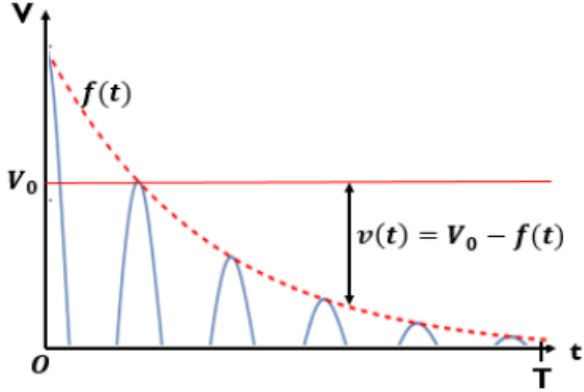

**Fig. 3** Diagram of DUT-Load resistor series.

Finding amplitude of weak signal requires much more effort because $V_{out}$ can gives only few limited information about $V_{signal}$. Still, it is not impossible if there are dozens of signal detector composed of Schmitt trigger and noise generator. Let the function representing amplitude of wave packet at time $t$ be $f(t)$ and the threshold of Schmitt trigger be $V_0$. Then the difference between these two value can be defined as $v(t) = V_0 - f(t)$. For given noise $V_{noise}$, transition occurs if $v(t) \leq V_{noise}$. Let $P_1(t)$, $P_2(t)$ be

$$\text{Probability of } v(t) \leq V_{noise} : P_1(t) \quad (2)$$
$$\text{Probability of } v(t) > V_{noise} : P_2(t)$$

Since the Gaussian white noise follows normal distribution shown in Eq.3 & 4, where $\Phi(x)$ is cumulative distribution function of normal distribution. Let $t_0$ be the time when last transition occurs. Probability of $t=t_0$ for some $t \in [0,T]$ is independent of past history ($T$ is time of measurement). Thus probability $t=t_0$ can be written as Eq. 5 where $t_0 + N \Delta t = T$. Taking ln both side gives Eq.6.

$$P_1(t) = \frac{1}{\sqrt{2\pi\sigma^2}} \int_{-\infty}^{v(t)} e^{-x^2/\sigma^2} dx = \frac{1}{\sqrt{2\pi}} \int_{-\infty}^{v(t)/\sigma} e^{-x^2} dx = \Phi\left(\frac{v(t)}{\sigma}\right) \quad (3)$$

$$P_2(t) = \frac{1}{\sqrt{2\pi\sigma^2}} \int_{v(t)}^{\infty} e^{-x^2/\sigma^2} dx = \frac{1}{\sqrt{2\pi}} \int_{v(t)/\sigma}^{\infty} e^{-x^2} dx = \Phi\left(-\frac{v(t)}{\sigma}\right) \quad (4)$$

$$P(t = t_0) = P_1(t_0) \prod_{n=1}^{N} P_2(t_0 + n \cdot \Delta t) \quad (5)$$

$$\ln P(t = t_0) = \ln P_1(t_0) + \sum_{n=1}^{N} \ln P_2(t_0 + n \cdot \Delta t)$$
$$= \ln P_1(t_0) + \sum_{n=1}^{N} \ln \Phi\left(\frac{v(t_0 + n \cdot \Delta t)}{\sigma}\right) \Delta n$$
$$\simeq \ln P_1(t_0) + \int_{n=1}^{N} \ln \Phi\left(\frac{v(t_0 + n \cdot \Delta t)}{\sigma}\right) dn \quad (6)$$
$$= \ln P_1(t_0) + \int_{t_0}^{T} \frac{1}{\Delta t} \ln \Phi\left(\frac{v(t)}{\sigma}\right) dt$$

Thus, the $P(t = t_0)$ becomes as shown in Eq.7. And finally, mean of $\langle t_0 \rangle$ can be calculated.

$$P(t = t_0) = P_1(t_0) \cdot \exp\left[\int_{t_0}^{T} \frac{1}{\Delta t} \ln \Phi\left(\frac{v(t)}{\sigma}\right) dt\right]$$
$$= \Phi\left(-\frac{v(t_0)}{\sigma}\right) \cdot \exp\left[\int_{t_0}^{T} \frac{1}{\Delta t} \ln \Phi\left(\frac{v(t)}{\sigma}\right) dt\right] \quad (7)$$

$$\langle t_0 \rangle = \int_0^T t_0 \cdot P(t = t_0) dt_0 \tag{8}$$

$$= \int_0^T t_0 \cdot \Phi\left(-\frac{v(t_0)}{\sigma}\right) \cdot exp\left[\int_{t_0}^T \frac{1}{\Delta t} \ln \Phi\left(\frac{v(t)}{\sigma}\right) dt\right] dt_0 \tag{9}$$

$$\begin{aligned} <t_0> &= F(\sigma, v(t)) \\ \Rightarrow <t_0> &= F(\sigma, b) \end{aligned} \tag{10}$$

This formula shows that $\langle t_0 \rangle$ is a function of $\sigma$ and $v(t)$ (or $f(t)$). And below of Eq.10 can be said if other values are fixed except for the decay constant. Thus, at least mathematically, amplitude of $V_{input}$ can be found by measuring $\langle t_0 \rangle$. To get $t_0$, dozens of detector is needed. For higher accuracy, fitting of $\langle t_0 \rangle - \sigma$ curve is required but this demands hundreds of detector (since we need dozens of detector for each $\sigma$). Fortunately, AD2 can reproduce the same signal several times, only one detector is sufficient in this case. Input signal was set as exponentially decaying sine signal with $A = 0.1V$, $f = 1000Hz$. SD of noise ranges from 0V to 0.5V by 0.01V and $V_{dc} = 4V$. Acquisition time and sampling rate was set as 1.5s, 20000Hz respectively. For each $\sigma$, 50 times of measurement were done to get $\langle t_0 \rangle$. To see the changes of $\langle t_0 \rangle - \sigma$ curve for different wave packet, input signals with $b = 1,3,5,7,9s^{-1}$ were tested. Proper nonlinear fitting was done on each curve and the coefficients of fit were compared for different $b$.

## III. RESULTS
### A. Implementation of bi-stable system: Schmitt trigger showing stochastic resonance
#### A.1. Transition of output signal

As shown in Fig.4, in the presence of noise with SD (Standard Deviation) value of 0.05V, the output voltage shows transition with respect to the sinusoidal input pulse. However, there are some shakes at both of output values. This can be interpreted as two ways. First, the op-amp used for our experiment isn't ideal. Also, the error from measurement by AD2 and LABVIEW program is inevitable. By installing more capacitors the output value will be shown clearer. But, the main point is that the result shows transition at appropriate point. When the input signal's displacement is near the positive maximum, the transition occurs from positive to negative. For the displacement being near the negative maximum, vice versa.

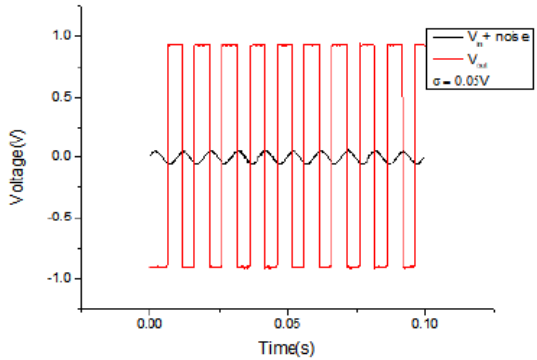

**Fig. 4** $V_{in}$, $V_{out}$ -time graph. Red solid line and black solid line is output signal and input signal + noise, respectively. The SD ($\sigma$) of noise is 0.05V. The amplitude of input signal is 0.05V, considering the voltage divider (the value we gave is 0.1V). Twice of threshold voltage is approximately ±0.09V from Eq.1

#### A.2. Signal to noise ratio

The SNR (signal to noise ratio) evaluated with respect to the values of noise SD is shown in Fig.5. For the noise sample rate of 4000Hz, the maximum SNR value is $38.1759 \pm 0.7145[1\sigma]dB$ at $\sigma = 0.055V$. For 20000Hz, the maximum SNR value is $48.6788 \pm 2.6960[1\sigma]dB$ at $\sigma = 0.05V$. In both case the SNR-$\sigma$ curve has local maximum value. And the curve shows known behavior. This gives us evidence of stochastic resonance with Fig.4 as mentioned in previous introduction section.

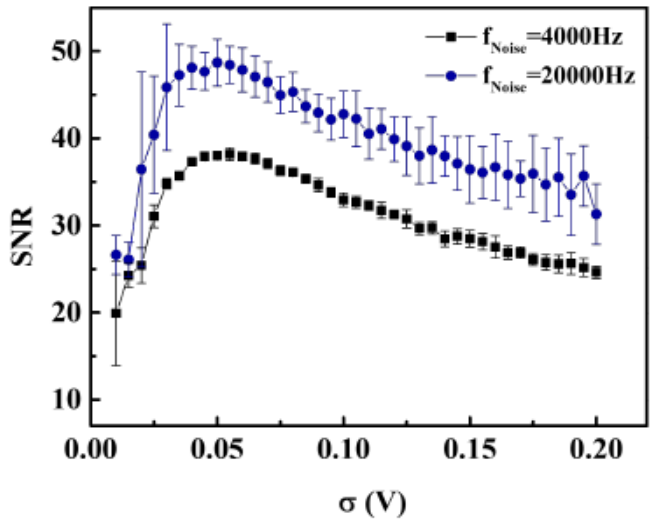

**Fig. 5** SNR(dB)-σ(V) curve. No non-linear fit is done since the actual known mathematical relation doesn't exist. The blue and black points are SNR values for noise sample rate of 20000Hz and 4000Hz, respectively. Peak points are (0.05V, 48.6788±2.6960[1σ]dB) and (0.055V, 38.1759±0.7145[1σ]dB) for 20000Hz and 4000Hz, respectively. The input signal is sinusoidal with frequency of 500Hz and amplitude of 0.05V, for both. The minimum and maximum value of noise is limited to -5 and 5V, respectively for both cases. And the twice of threshold voltage of the system is approximately ±0.09V from Eq.1, also for both cases.

### A.3. Hysteresis of Schmitt trigger

The hysteresis graph is obtained as shown in Fig.6. With increasing the input signal's voltage, at $V_{in} = -0.075$V, the transition of output voltage value is shown. Blue line represents it. By decreasing the $V_{in}$, at $V_{in} = 0.098$V, the transition occurred. This well fits with characteristic of Schmitt trigger system with threshold in it like Fig.4. And This hysteresis gives good validation of the system's bi-stability as mentioned in the previous. With comparison of twice of threshold voltage value: ±0.09V, experimentally obtained twice of upper and lower threshold values are -0.075 & 0.098V, respectively. Thinking of absolute value, it has slight difference from threshold voltage obtained from Eq.1. This is because of op-amp's non ideal characteristic. The equation is earned in assumption of ideal op-amp, however in reality it is not. So it can be concluded that hysteresis gives us validation for treating the Schmitt trigger as a bi-stable system with two different threshold voltages of -0.0375 & 0.049V. Fig.7 is predicted physical model of Schmitt trigger.

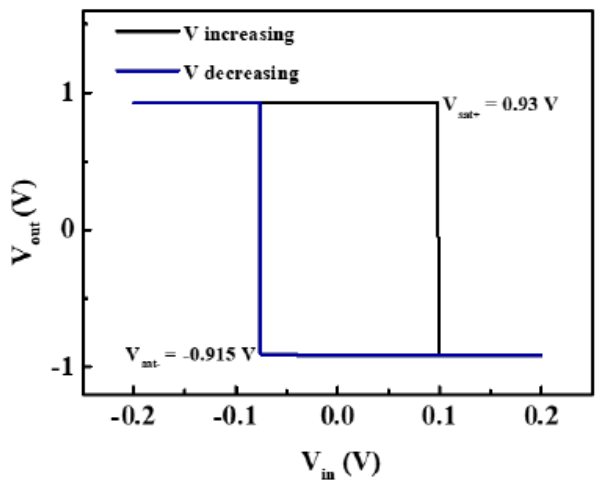

**Fig. 6** Hysteresis loop in terms of $V_{out}$ and $V_{in}$. The blue and black solid lines have direction of decreasing (positive to negative sign) and increasing (negative to positive sign), respectively. Saturated output voltage is 0.93 and -0.915V for increasing and decreasing case. The input signal is DC signal in this case and no noise is generated.

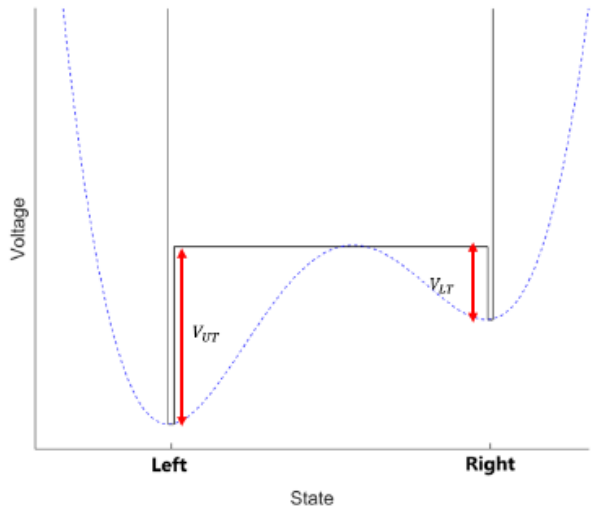

**Fig. 7** Bi-stable system made with Schmitt trigger circuits. The black solid line represents Schmitt trigger system. Only two states exist. Left represents one of state and if from this state, it overcomes threshold of 0.049V, transition towards Right occurs. By overcoming threshold of -0.0375 (the input signal + noise is bigger than absolute value 0.0375.), opposite transition occurs. Blue dashed lined is shape of generally known bi-stable systems. The difference is that the stable state's width is narrower for Schmitt trigger.

### A.4. Relation between threshold and DC power

As now from result of hysteresis loop, real threshold values in terms of giving ±1V as DC power are obtained with respect to the value 0.09/2 = 0.045V which is obtained from Eq.1. So for further information about threshold voltages of our bi-stable system, relation between DC power and the threshold voltage is measured. Fig.8 shows linearity of threshold voltage with respect to the DC power. $V_{th} = 0.051V_{dc} - 0.005$ is the formula from the linear regression. Following this, in terms of DC power of 1V, the threshold voltage value is 0.046V which has difference of 0.001V with 0.045V.

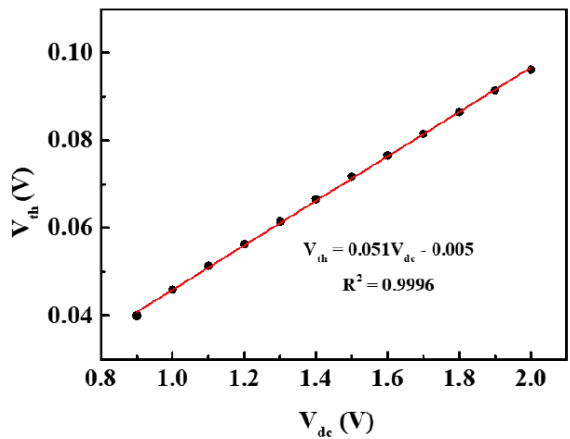

**Fig. 8** Graph of $V_{th}$-$V_{dc}$. Linear regression with Origin program is done. The proportional coefficient is 0.051 and the offset is -0.005. $R^2$ value is 0.9996.

## B. Weak signal detection: frequency of damping sinusoidal wave
### B.1. Measured frequencies of input signal

Following the method from previous section-method, the result of FFT is shown in Fig.9. Since the result of FFT gives frequency's occupancy of output signal, the occupancy which in this case the logarithm scale of amplitude RMS is highest in near 0Hz. Because the input signal is damping sinusoidal wave which causes pause in transition of output signal eventually or in same words, 0Hz of output signal frequency. So by finding second highest peak the frequency values are obtained. As shown in Fig.9, the peak gets narrower as the input signal's frequency increases. For input frequency of 10Hz, the peak is not found. Table.1 shows the error rate of obtained frequency with respect to the real value. As the given frequency increases, the error rate is lessened.

**Table 1** The error rate of obtained frequency with respect to the given frequency.

| Given frequency[Hz] | Error[%] |
|---|---|
| 50 | 14 |
| 100 | 2.3 |
| 500 | 0.14 |
| 1000 | 0.04 |
| 2000 | 0.04 |

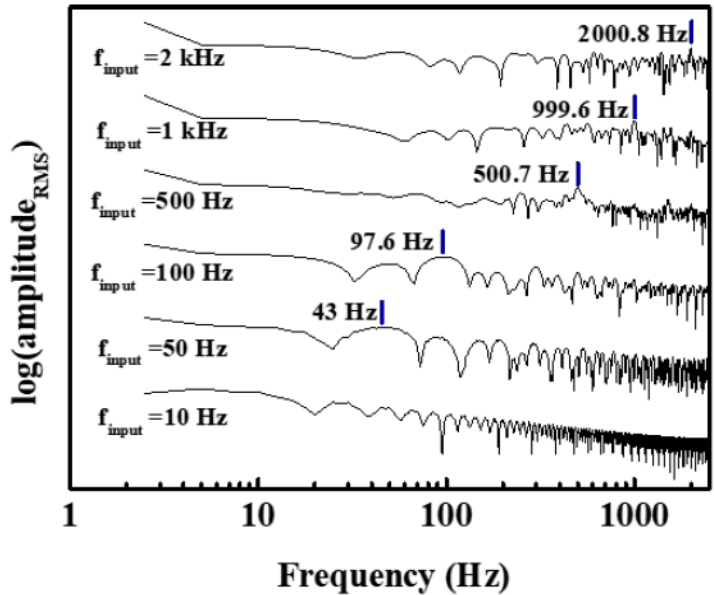

**Fig. 9** FFT result with graph of log(amplitude RMS)-frequency. The input signal's frequency values are written in the plot. The blue solid line shows the position of second highest peak with obtained frequency values on the left side of the line. For frequency of 10Hz, the peak is not detected. The amplitude of damping sinusoidal signal is 0.1V with decay constant of 5[$s^{-1}$]. SD of noise is 0.01V. And the DC power supply was set as 1V which causes threshold voltage of 0.046V from Fig.8.

**B.2. Limitation of frequency detection**

From the previous section, it is shown that in low frequency area (briefly smaller than 100Hz), the FFT results show smeared peaks. In this term, using MATLAB pro gram, Fig.10 is obtained. It shows that in high frequency area the error rate is nearly 0 which implies that detection of frequency is reliable. However, for low frequency area the error rates are on a scale of 10%. In short, the method of using stochastic resonance for detecting the unknown frequency might have limitation in low frequency area.

**B.3. Complementary for the limitation**

Again, since the FFT gives us occupancy of frequency, it is needed to make the occupancy of output signal's frequency (which fits with the input signal's unknown frequency) higher. At first, it is intuitive to think that by giving higher SD of noise the duration of stochastic resonance increases. And further, the occupancy of the output signal's frequency increases. As shown in Fig.11, with comparison of $\sigma = 0.01V$, 0.02V and 0.03V, the 0.03V one shows least error rate 0.28%. with respect to 50Hz. However, comparing the $\sigma = 0.03V$ and 0.05V, it can't be just concluded that error rate decreases as SD of noise increases. So the SNR-$\sigma$ curve is obtained as shown in Fig.12. It is shown that $\sigma = 0.03V$ is the point of local maximum. SNR is supposed to be the index of how well the output signal is detected in presence of noise. So it is quite well explained if we think that $\sigma$ has its optimal value for detecting the output signal's frequency and thus detecting the input signal's frequency. And the optimal value can be evaluated by obtaining SNR curve.

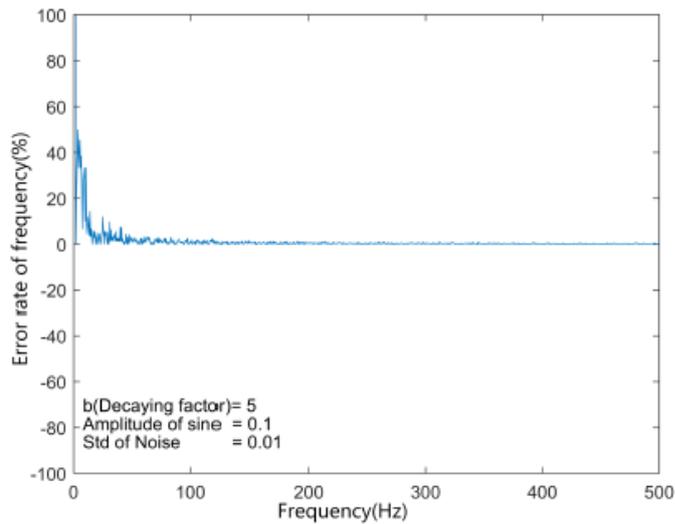

**Fig. 10** Simulated result of error rate-frequency. The frequency is what we gave for the input signal. Amplitude of damping sinusoidal signal is fixed as 0.1V with the decaying factor is 5[s$^{-1}$]. The frequency of input signal is from 0 to 500Hz with increments of 0.5Hz.

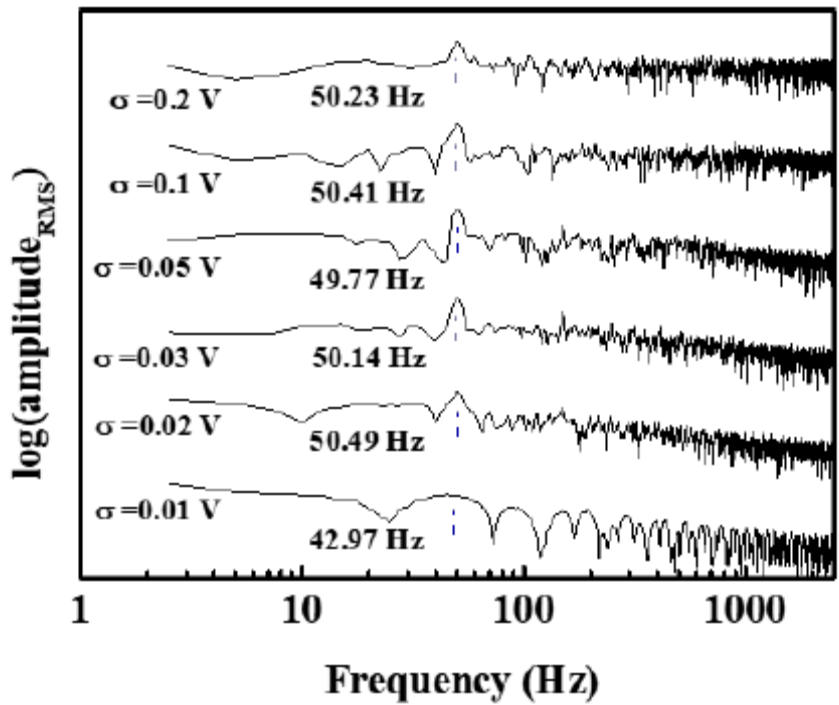

**Fig. 11** FFT results at low frequency area. The input signal's frequency is 50Hz with amplitude of 0.1V and decay constant of 5[s$^{-1}$]. The dashed line shows the obtained frequency.

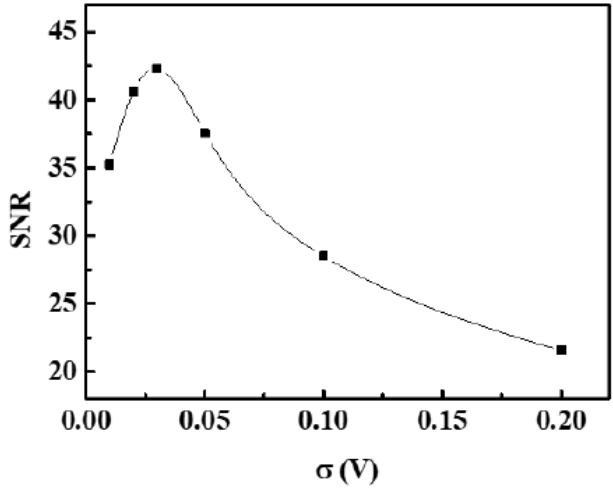

**Fig. 12** SNR curve with respect to $\sigma$ values from Fig.11. The local maximum is shown in $\sigma = 0.03$V.

**C. Weak signal detection: amplitude of damping sinusoidal wave**

Following the method from previous section-method, the relations of $\langle t_0 \rangle$ and SD of noise with constant decay constant are shown in Fig.13. For different decay constant 1 to 9[s$^{-1}$] with increments of 2[s$^{-1}$], five graphs look similar to each other. With trial of several non-linear fit, the sigmoid function has shown the highest value in R$^2$. Therefore, according to Eq.10, it is possible to claim that $\langle t_0 \rangle$ and $\sigma$ has relation following sigmoid function, approximately. The sigmoid function fit was done with form of Eq.11. Two unknown parameter $A$ and $B$ are included in it. Fig.14 & 15 shows five evaluated values of $A$ and $B$ by change of decay constant. Both parameter $A$ and $B$ show a certain tendency to increase as decay constant does. This is quite easy to understand with reference of Eq.10. Since $\langle t_0 \rangle$ is function of two variables-SD of noise & decay constant-if the other elements of input signal are fixed as constant. Eq.11 gives $\langle t_0 \rangle$ as a function of $\sigma$. And the parameters $A$ and $B$ might be functions of decay constant in this term of interpretation. In short, $\langle t_0 \rangle$ is composite function of $\sigma$ and decay constant with high probability as shown in Eq.12. Since Eq.9 can't be solved analytically, Fig.14 & 15 aren't fitted with non-linear curves. If Eq.9 can be solved with numerical approximation, the form of Eq.12 might be clearer and thus, the relation between parameter $A$, $B$ and decay constant can be obtained. If this can be done, by measuring $\langle t_0 \rangle$ from different detectors of different SD values of noise, parameters $A$ and $B$ can be evaluated and with reference of their relation with decay constant, the damping of input signal can be obtained. Further, it will give us information about envelope and thus, amplitude also.

$$<t_0> = \frac{1.5}{1+exp[-A(\sigma - B)]} \quad (11)$$

$$<t_0> = F(\sigma) \circ A(b) \circ B(b) \quad (12)$$

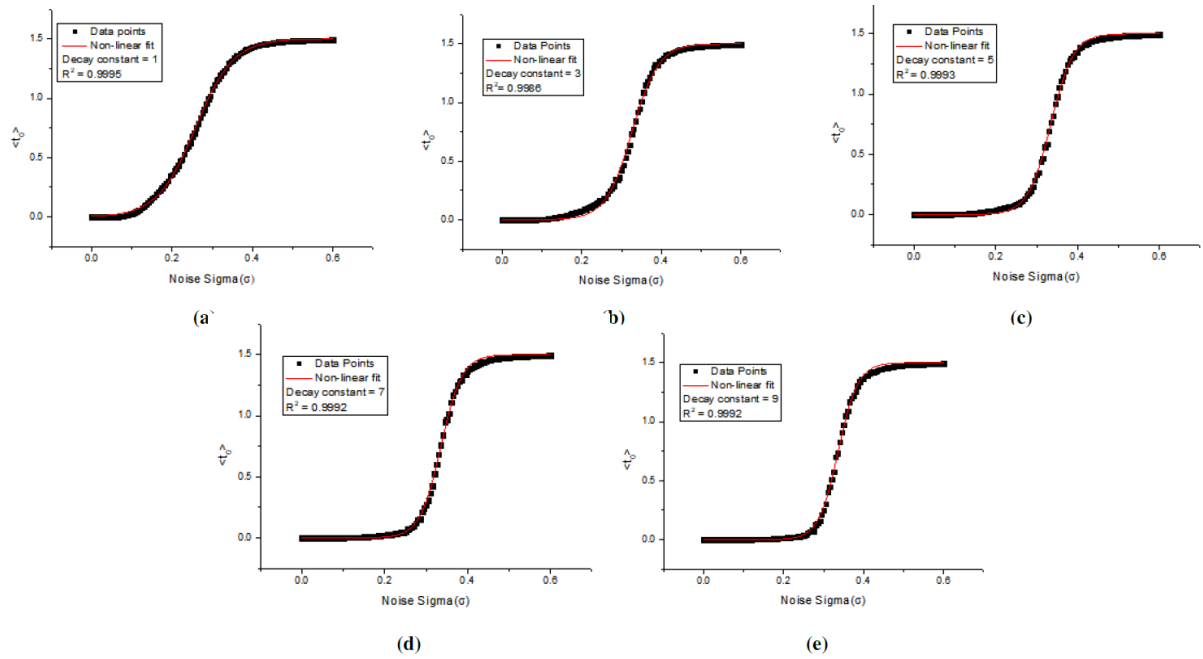

**Fig. 13** Graph of ⟨$t_0$⟩-$\sigma$. The damping signal's amplitude and frequency was set as 0.1V and 1000Hz, respectively. SD of noise ranges from 0 to 0.5V with increments of 0.01V and the power supply was set as 4V which makes the threshold voltage $V_{th}$ = 0.199V from Fig.8. Data points are fitted with non-linear curve fit, especially sigmoid function. (a): Decay constant is 1[$s^{-1}$] and $R^2$ = 0.9995. (b): Decay constant is 3[$s^{-1}$] and $R^2$ = 0.9986. (c): Decay constant is 5[$s^{-1}$] and $R^2$ = 0.9993. (d): Decay constant is 7[$s^{-1}$] and $R^2$ = 0.9992. (e): Decay constant is 9[$s^{-1}$] and $R^2$ = 0.9992.

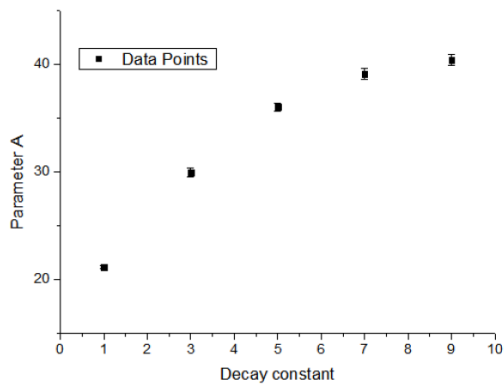

**Fig. 14** Plot of parameter *A* with respect to the decay constant value.

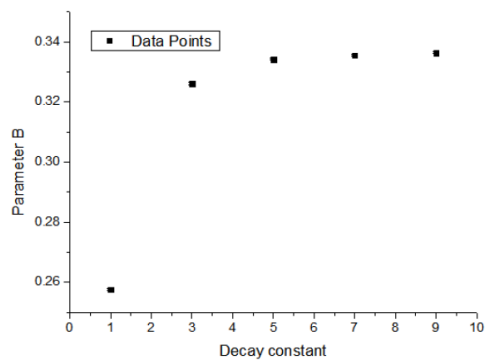

**Fig. 15** Plot of parameter *B* with respect to the decay constant value.

## IV. DISCUSSION
### A. Basic Method of Detection

As shown in Fig.16, set of detectors are needed for weak signal detection in our case. Each detector implies a bi-stable system we constructed. These bi-stable systems are identical except for the noise. We generate artificial noise to each bi-stable system varying its SD. All the detectors have same threshold, which is set with sufficient range with respect to the arbitrary input signals' amplitudes. So only few detectors will show stochastic resonance. For the frequency detection, by analyzing the output signals from detectors which show stochastic resonance, the detection can be done. Using FFT as shown in the previous section, the output signal's frequency will be obtained. This will be the weak signal's unknown frequency. And it is well verified with our results. Even though the frequency detection has limitation in low frequency area, complementary can be done. Because the detectors differ in SD of noise, there must be some noises with optimal SD values. This implies us that lots of detectors should be needed since varying the SD in smaller increments will give us more precise information. Now, for the detection of amplitude, in presence of damping input signal, only some detectors will show stochastic resonance. The systems will vary only in SD of noise. And eventually the resonance will stop at different values of $t_0$ for each detector. And for a single detector that has shown stochastic resonance, $t_0$ can be measured for several times. Eventually by obtaining $\langle t_0 \rangle$ from each detector, $\langle t_0 \rangle - \sigma$ curve can be obtained. And in the assumption that numerical approximation of Eq.9 is possible, decay constant can be obtained. Which gives us information about envelope and amplitude. This was about amplitude of damping input signal. What about a signal with no damping? In this case, $t_0$ which is a time point that transition (of output signal) stops is not defined well. Because the bi-stable system with SD over a certain value will show stochastic resonance permanently. So in this case another type of set is needed. This time briefly thinking, the detectors differ in threshold and have same noise (same SD) generated. Say an input signal is sine pulse with amplitude of 0.01V. And the thresholds vary from 0.001V to 0.1V. If the SD we set is not that huge with respect to 0.1V, not all detectors will show the stochastic resonance. In other words, the transition of output signal will be shown in low threshold detectors and there must be a detector whose threshold is big enough so that no stochastic resonance occurs. That threshold will be the reference of amplitude of the sine pulse. So this method is the way using calculation of unknown amplitude's range. However, in reality we don't have any information about weak signals. So the ideal method would be this. Prepare sets of detectors. For detectors in one set differ only in SD of noise. And each set differ only in the threshold of bi-stable system. The smaller the differences of SD & threshold are, the less the frequency & amplitude errors are.

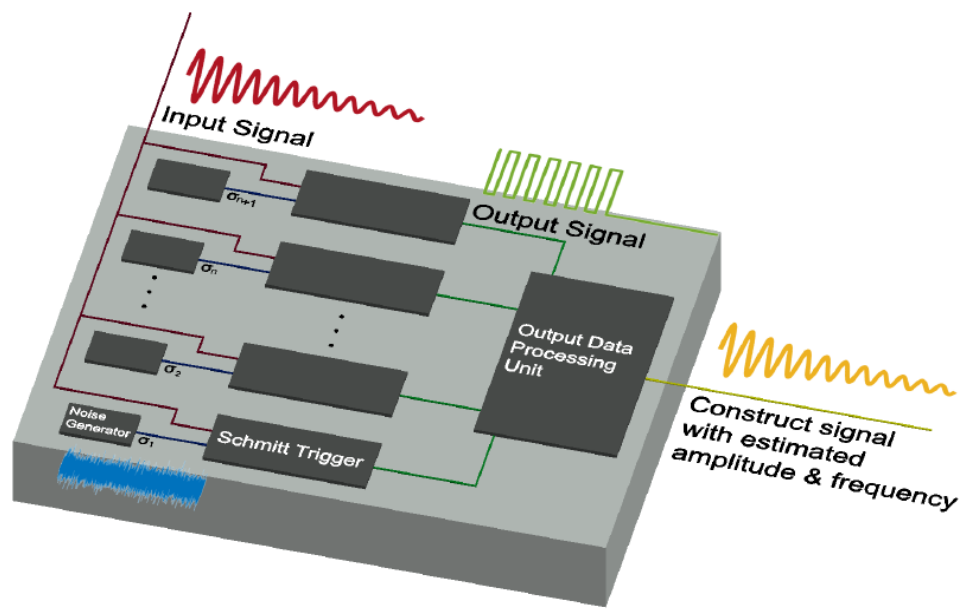

**Fig. 16** Schematic of set of detectors made of parallel Schmitt trigger circuits.

## B. Quantum stochastic resonance

There are some results of stochastic resonance shown in quantum systems[8]. By implementing quantum dots in semi-conductor the dot itself can be bi-stable quantum system. Quantum dot is a hole with quantized energy levels inside. With setting somewhat called gate, in middle of the electron reservoirs a hole or dot can be made. At the reservoir, the electrons can be filled just below the Fermi level. Between the dot and the reservoirs or ohmic contacts, potential barriers exist. And by tunneling electrons can move into or out from the dot. In this respect, if the gate voltages which determine the barrier ($V_{d1}$ & $V_{d3}$) and the gate voltage which determine the energy level of quantum dot ($V_{d2}$) are well synchronized with the tunneling rate or in other words, tunnel coupling between the ohmic contacts and the dot, the sequence of electron transportation can occur. This causes the relative current of QPC-detector to be changed with respect. And the output signal in this quantum system is the current signal of QPC.

### B.1. Stochastic resonance in the quantum dot system: in terms of 3 components

In the paper "Quantum stochastic resonance in an a.c.-driven single-electron quantum dot"[8], the main theme they are willing to talk about is discovery of stochastic resonance. Or more precisely, quantum stochastic resonance. As we know, stochastic resonance needs 3 big components. Weak signal, noise and bi-stable system. In general, stochastic resonance studied in electric engineering fields[1]~[5] is shown by applying artificial noise to a weak signal (it is already mixed with noise naturally generated) and make the state of the system change periodically by going over the threshold. However, in the paper, the noise is thought to be as intrinsic component of the physical system. In other words, tunneling (precisely, tunnel coupling) is regarded as intrinsic noise which is synchronized to the weak signal and changes the state of the system. So there is no need to generate artificial noise in this case. Also, the weak signal in this case is not the unknown signal. The controlled gate voltages can be regarded as the weak signal. Since the electron itself can be two-level system, vaguely interpreting, we can think the dot itself is a bi-stable system. However, the output pulse is current of the QPC which is determined by the occupancy of electron in the dot. So it can be interpreted that bi-stable system in this case consists of two states of electron is in the dot or not. And the threshold would be tunneling barriers between the dot and the reservoirs (ohmic contacts). In short, in the presence of tunneling (intrinsic noise), gate voltage (weak signal) is applied. And if the tunneling barrier (threshold) is able to be crossed, electron can get out or get in and stochastic resonance can be shown.

### B.2. Method of detection using the quantum dot system

The basic concept of detection will be almost identical from the previous method. However, the fact that should be mentioned is with using this system, the unknown weak signal should be received and act as a gate voltage. So for this method the set of detectors or in other words, systems should be prepared varying in tunnel coupling (intrinsic noise) so that only few synchronizes with the gate voltage and show the stochastic resonance. The detection of frequency will be almost identical with the previous one. However, for amplitude detection this method might be easier. Assume that one detector showed stochastic resonance which means tunneling in and out with respect to the dot occurs stochastically. Then it can be indicated as there is transition between two states- electron occupancy yes/no. If we can measure the transition probability with counting statistics of electrons, then it is possible to know the amplitude. For example, if the perturbation is sinusoidal Eq.13[9] is satisfied where $H'$ is perturbation (weak signal) and $P_{a \to b}$ is the probability of transition between two states $a$ and $b$. $\omega_0$ is the transition frequency. So if we can measure the probability and know all the factors, $V_{ab}$: the unknown amplitude can be obtained. Since quantum systems are highly sensitive to perturbations, weak signals that can't be seen with classical detection methods would be possibly detected. In this sense, this method can have high potential in detection field.

$$H'_{ab} = V_{ab} cos(\omega t)$$
$$P_{a \to b} \approx \frac{|V_{ab}|^2}{\hbar^2} \frac{sin^2[(\omega_0 - \omega)t/2]}{(\omega_0 - \omega)^2} \quad (13)$$

## V. CONCLUSION

In our research, mainly using conventional circuit elements such as op-amp, Schmitt trigger was implemented in order to represent bi-stable system. Several measurements were done for validation of bi-stability and stochastic resonance. The transition of output signal and SNR curve would be good proofs of stochastic resonance. Furthermore, the hysteresis implies the bi-stability of our implemented system. After verifying well-done implementation, we constructed two detection experiments-frequency & amplitude of damping sinusoidal pulse. It is shown that frequency detecting has limitation in low frequency area. However, the fact that complementary with proper SD of noise can be possible is also verified. For the amplitude detection, we got stuck on mathematical problems. But, by identifying some relations or tendency of measured values, it is concluded that in presence of proper numerical approximation, amplitude can also be detected. The further discussion gives us how detection using stochastic resonance should be constructed. Also it is discussed that in terms of quantum stochastic resonance, detection method suggested in this report can have higher potential. In short, by trimming the method, stochastic resonance really is applicable for detection of weak signal.